\documentstyle[aps]{revtex}

\begin{document}
\input psfig

\title{The Worldsheet Formulation as an Alternative Method 
for Simulating Dynamical Fermions.}

\author{Hugo Fort \\
Instituto de F\'{\i}sica, Facultad de Ciencias, \\
Tristan Narvaja 1674, 11200 Montevideo, Uruguay}

\maketitle

\begin{abstract}
The recently proposed worldsheet formulation of lattice
fermions is tested for the first time carrying out 
a simulation for the simplest model: the one-flavor, strictly
massless lattice Schwinger model. 
A main advantage of this alternative method for
simulating dynamical fermions consists in its economy:
it involves many fewer degrees of freedom than the
ordinary Kogut-Susskind formulation. 
The known continuum limit is reproduced by the 
method for relatively small lattices.
\end{abstract}

\vspace{1mm}


It is well known that a complete description of Quantum 
Chromodynamics (QCD) requires a 
non-perturbative approach. 
The main available non-perturbative 
method is lattice gauge theory. In particular, the {\em Lagrangian 
formulation}, with discretized Euclidean time, allows to 
apply the powerful numerical method of Monte Carlo simulation. 
However, simulating  QCD
with dynamical fermions or {\em full} QCD 
is still too expensive in computer time.
Throuhout the years the conclusion has been the same: 
either more machine power and/or a
real improvement in algorithms
is needed to produce reliable estimates of hadronic 
quantities \cite{lat}.

The root of the problems of QCD can be traced to
the fact that the theory is defined in terms of 
local fields, quarks and gluons, but
the physical excitations are extended composites: 
mesons and baryons. A Hamiltonian
lattice formulation directly in terms of string-like 
excitations gave rise to the so-called 
{\em P-representation} \cite{fg},\cite{gs}.
The states $|P>$ of this representation are described by 
sets of lattice paths associated to the extended
excitations.
We propose to explore a new numerical approach 
based on the recently introduced worldsheet formulation 
of lattice gauge theories with dynamical fermions \cite{afg}.
This formulation, in terms of 
the worldsheets of the extended excitations, is the 
Lagrangian counterpart of the P-representation.
The worldsheet partition function of lattice QED with 
staggered fermions is expressed as a sum over 
surfaces with border on self-avoiding fermionic loops.
Each surface is weighted with a classical action 
written in terms of integer gauge invariant variables.
The surfaces correspond to the worldsheets of 
loop-like pure electric flux excitations and meson-like
configurations (open electric flux tubes 
carrying matter fields at their ends). 
This description, besides the general advantage of 
geometric transparency,
is appealing because 
it involves fewer degrees of freedom than the
ordinary Kogut-Susskind formulation implying  
a substantial economy of CPU time.
Furthermore,
the present formulation of dynamical fermions in terms
of self-avoiding loops is closely connected with 
the polymer-like representation 
\cite{mon} and this might enable  
to exploit techniques used in other branches of physics,
like condensed matter physics \cite{wie} and cosmic string
physics \cite{cop}.

In this letter we present, for the first time,
a test of the formalism developed in \cite{afg}.
The simplest lattice 
gauge theory with dynamical
fermions, the Schwinger model or (1+1) QED, is chosen
and  a Monte Carlo simulation is performed.  
This massless model can be exactly solved in the continuum.
However, it is rich enough to share many features 
with 4-dimensional QCD. 
For this reason it has been extensively used as a laboratory to 
analyze the previous phenomena and has also become a popular
benchmark to test different techniques to simulate
theories with fermions. 
Our main point is to show that the method works properly
and it might provide a novel alternative approach to 
simulate {\em full} QCD.

        The P-representation offers a 
gauge invariant description of physical
states in terms of kets $\mid P >$, 
where $P$ labels a set of connected paths
$P_x^y$ oriented from the even sites $x$ to the odd
ones $y$ in a lattice of spacing $a$.  
These paths correspond to string-like ``electromeson''
excitations of staggered fermions  
connected by tubes of electric flux. 
The internal product of a ket $\mid P_x^y >$ with one 
in terms of fields is given by 
\begin{equation}
\Phi (P_x^y) \equiv <P_{x}^{y} \mid {\psi}_u^{\dagger} ,
{\psi}_d, \theta_{\mu} > 
= {\psi}_{u}^{\dagger} (x) U(P_x^y) {\psi}_{d}(y),
\label{eq:Phi}
\end{equation}
where $U_\mu(x)\equiv
\exp [i\theta_\mu(x)]\equiv \exp [iea A_\mu(x)]$ and
$u$ and $d$ denote respectively the up and down parts
of the Dirac spinor.
Therefore, the lattice 
paths $P_x^y$ start in
sites $x$ of a given parity and end in sites 
$y$ with  opposite parity.
The one spinor component at each site 
can be described in terms of the Susskind's 
$\chi (x)$ single Grassmann 
fields \cite{sus}. The path creation  
operator $\hat{\Phi}_Q$ in the space of kets $\{ |\, P> \}$
of a path with ends $x$ and $y$ is defined as 
\begin{equation}
\hat{\Phi}_Q= \hat{\chi}^{\dagger} (x) \hat{U}(Q_x^y) 
\hat{\chi}(y).
\label{eq:Phiop}
\end{equation}
Its adjoint operator $\hat{\Phi}_Q^{\dagger}$ acts
in two possible ways \cite{fg}: annihilating the path $Q_x^y$
or joining two existing paths in $|\, P>$ one ending at
$x$ and the other starting at $y$.

The worldsheet partition function $Z_P$ is a sum over 
the worldsheets of strings or paths of the P-representation. 
That is, surfaces $S_{{\cal F}^c}$ such that: {\bf (I)} their borders 
${\cal F}^c$ are self-avoiding polymer-like loops and {\bf (II)}
when intersected with a 
time $t=\mbox{constant}$ plane they produce
paths beginning at even sites and ending at odd ones. 
This description at first sight is similar to the one obtained
by integrating the fields in the Kogut-Susskind partition
function $Z_{KS}$, nevertheless
it differs in two features. In first place, the integration
of the fields in $Z_{KS}$
produces, besides surfaces bounded by fermionic loops,
isolated links traversed 
in both opposite directions or ``null" links. These
``null" links were ruled out from the $Z_P$ and with them 
the myriad of different configurations
corresponding to a given configuration of worldsheets.
In second place, by virtue of the constraint {\bf (II)},
the surfaces of $Z_P$ when intersected with a 
time $t=\mbox{constant}$ plane produce only
paths beginning at even sites and ending at odd ones
instead of the paths having ends of any parity (as it
happens with the surfaces obtained from $Z_{KS}$).
These two differences translate respectively in: computer
time economy and a cure for the additional
species doubling problem of the Kogut-Susskind action.
Concerning the last point, 
the worldsheet action has only 2 fermion species
in 4 dimensions (instead of 4 fermion species) which is
phenomenologically more satisfactory. 
In ref.\cite{afg} it was proved that $Z_P$ leads to 
the QED Hamiltonian using the transfer matrix procedure. 
The expression for $Z_P$ is as follows \cite{foot1}:
\begin{equation}
Z_P= 
\sum_{S_{{\cal F}^c}} \, 
\sigma({\cal F}^c) \, \exp\{
-\frac{1}{2\beta} \sum_{p \in S_{{\cal F}^c}} n_p^2
\},
\label{eq:ZP0}
\end{equation}
where $n_p$ is an integer variable attached to plaquettes 
(a 2-form) and $\sigma_{\cal F}$ is a sign 
given in terms of purely 
geometric quantities of the fermionic loops ${\cal F}^c$ \cite{afg}.
In $D=2$, it turns out that 
$\sigma({\cal F}^c)=(-1)^{N_{{\cal F}^c}-
\frac{L_{{\cal F}^c}}{2}+A_{{\cal F}^c}}$ 
-- where $N_{{\cal F}^c}$, $L_{{\cal F}^c}$ and
$A_{{\cal F}^c}$ are, respectively, the number of connected
parts, the length and the area of ${\cal F}^c$ --  
and all the 
non-vanishing contributions have $\sigma_{\cal F}=+1$.
The reason is that $N_{{\cal F}^c}-
\frac{L_{{\cal F}^c}}{2}+A_{{\cal F}^c}=I_{\cal F}^c$,
the number of enclosed sites by the fermionic 
loops ${\cal F}^c$ which is always even by virtue of
the above constraints {\bf (I)} and {\bf (II)} 
( see ref.\cite{afg} for more details), 
so that we omit this factor.

The fermionic paths ${\cal F}^c$ can be expressed in terms
of integer 1-forms
--attached to the links-- $f$ with three possible values: 0 and
$\pm$ 1 with the constraint that they are 
non self-crossing and closed ($\partial f = 0$) as
\begin{equation}
Z_P^{\mbox{\footnotesize Schwinger}}= 
\sum_{n} \, \sum_{f} \, 
\exp \{
-\frac{1}{2\beta} \|n\|^2
\}
\delta (f-\partial n).
\label{eq:ZP}
\end{equation}
The lattice chiral condensate per-lattice-site is defined 
as $<\! \bar{\chi}\chi \! >
=\frac{1}{2N_s}\sum_x (-1)^{x_1}
<[\hat{\chi}^\dagger 
(\mbox{x}),\hat{\chi}(\mbox{ x})]>$, where 
$N_s$ is the number of lattice sites.
The corresponding operator is realized in the 
P-representation and thus we get for 
the chiral condensate \cite{fg}:
\begin{equation}
<\bar{\chi}\chi >=
\frac{1}{2}-\frac{2{\cal N}_P}{N_s},
\label{eq:chir-cond2}
\end{equation}
where ${\cal N}_P$ is the number of connected paths at a given
time $t$.
Thus, equation (\ref{eq:chir-cond2}) allows to calculate directly
the chiral condensate simply by counting the number of 
``electromesons" we have when we intersect their world sheets
with each time slice $t$. 


To generate the worldsheets 
we use a Metropolis-type updating algorithm
with the Boltzmann weight  
proportional to $\exp \{-\frac{1}{2\beta} \sum_p n_p^2\}$.
We simulate the model on $L\times L$ square lattices with periodic 
boundary conditions ($pbc$).
All the plaquettes $p$ belonging
to a surface, by virtue of the self-avoiding 
constraint  {\bf (I)} on their borders, must fulfill
the condition that the difference between the integers $n_p$
of contiguous plaquettes has 3 possibilities: $\Delta_\mu n_p=\pm 1$
or 0 ($\mu = 1,2$). 
On the other hand, the constraint {\bf (II)} implies
that $\Delta_1 n_p=1 \mbox{or} -1$ according to the parity of
the spatial coordinate $x_1$. With these simple rules 
the algorithm generates the surfaces.
In the computation of the
chiral condensate per-lattice-site
$<\bar{\chi}\chi >$ there is a clear difference between 
a simulation in 
the ordinary representation, in terms of 
fields, and one in the P-representation.
In the first case, for the massless model, given 
enough time, the system
rotates through all the degenerate minima so that
$<\bar{\chi} \chi \,> =0$. Therefore, one has to calculate 
this order parameter for several small masses $m \neq 0$
, which select the $\theta=0$ vacuum, and then extrapolate
to the limit $m \rightarrow 0$. On the
other hand, in the gauge invariant P-representation
from the very beginning $\theta=0$ \cite{foot2} and the computation with
$m=0$ can be performed directly.
To compute $<\bar{\chi}\chi >$
using (\ref{eq:chir-cond2}) we have to count 
${\cal N}_P$ at each one of the $L$ time slices.
Therefore, for each lattice sweep we collect $L$ 
values of $<\bar{\chi}\chi >$. 
Typically, 100000 sweeps per point were performed.

The first thing we have checked is that we get the right strong
coupling behavior, both for the 
ground-state energy density $\omega_0$ and the chiral condensate 
per-lattice-site $<\bar{\chi}\chi >$. The agreement with the series 
expansion is pretty good up to $\beta=0.12$.

On the other hand, the continuum theory is reached 
at zero lattice coupling, in the same way as 
four-dimensional asymptotically
free theories like QCD.
It is known  that the chiral condensate in the 
continuum is given exactly by  
\begin{equation}
\frac{<\bar{\psi} \psi>}{e} = \frac{e^\gamma}{2\pi^{3/2}}=0.15995
\;\;\;\;\;\;\;(\gamma \mbox{ is Euler's constant} ). 
\label{eq:chiral-cont}
\end{equation}
As  the lattice size $L$ increase,
the convergence of $q\equiv <\bar{\chi}\chi >\!\!/\!e$ to 
its known continuum value (\ref{eq:chiral-cont}) improves. 
A property of the action of (\ref{eq:ZP}) is that 
$q$ becomes stabilized at its continuum
value well inside the weak coupling. Before reaching this
regime, this observable oscillates strongly with $\beta$. 
In FIG.1 we plot $q$ vs. $\beta$ 
for lattices of sizes ranging from $L=20$ to
$L=32$. For $L=32$ it is apparent that a value 
close to the exact
continuum one is reached. 

\begin{center}
\begin{figure}[t]
\hskip -3mm \psfig{figure=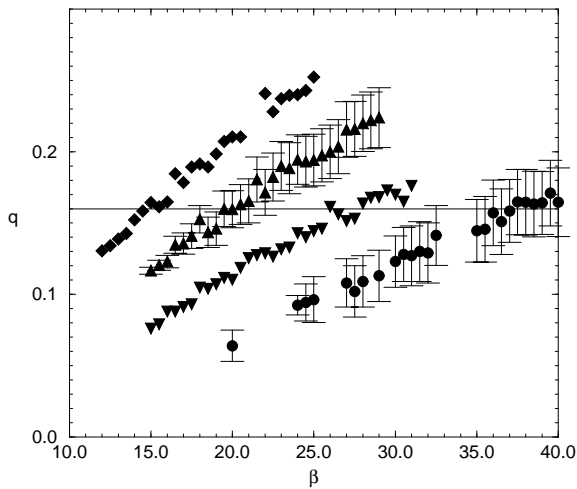,height=7.5cm}
\caption{$q\equiv <\bar{\chi}\chi >\!\!/\!e \,$
for different lattice
sizes $L=20$ (diamonds), $L=24$ (triangles up)
, $L=28$ (triangles down) and $L=32$ 
(circles). The horizontal line
correspond to the known continuum value $\simeq 0.16$ 
(statistical errors are indicated by vertical bars for $L=24$
and $L=32$). }
\label{fig1}
\end{figure}
\end{center}


Our aim has been to show that the recently introduced
worldsheet formulation is a valuable alternative tool
in order to do numerical
computations with dynamical fermions. So, at this stage,
we have chosen the simplest algorithm: 
the standard Metropolis one. Obviously, there are several more 
sophisticated techniques that can be applied in order to 
improve the results.
Additionally, the code can be, of course, optimized.

The method presents the following advantages: {\bf 1)} 
Easiness of computation. For instance,
the lattice chiral condensate is
diagonal in the P-representation and hence to evaluate
it one simply has to count the number of connected
open string-like excitations. Furthermore, 
one can compute
directly $<\bar{\chi}\chi >$ for the case of {\em strictly
0 mass}.  
{\bf 2)} Economy I: it involves 
much less degrees of freedom that the Kogut-Susskind action.
{\bf 3)} Economy II: no gauge redundancy. 
{\bf 4) } As a consequence of the constraint {\bf (II)}, 
the worldsheet action does not suffer from  
the additional species doubling problem of the 
Kogut-Susskind action.

Concerning the results: we get the right 
continuum limit (the weak coupling fixed point) as one can see
from the chiral condensate which is consistent 
with the exact results for the continuum theory.
This is a numerical confirmation of the redundancy 
of the information carried out by the ``null" links
and the correctness
of their elimination from the partition function.     

Finally, we would like to stress (once more) 
that our aim was not to present another 
solution to the Schwinger model, but to test a new general 
approach to tackle dynamical fermions.
The results are promising, and this encourages us to 
employ more refined numerical methods in a following stage.
To extend the method to $D > 2$
one has to compute
the sign $\sigma$ of each configuration. We are working on
an algorithm to compute efficently this factor. 

\vspace{10mm}

{\large \bf Acknowledgements}

\vspace{2mm}

I am indebted with J. M. Aroca and R. Gambini 
for many valuable discussions.
I also would like to thank V. Azcoiti, M. Baig and  G. Beltrame
for helpful technical comments.

\vspace{2mm}

This work was supported in part by CONICYT, 
Project No. 318.

\end{document}